\documentclass[twocolumn,preprintnumbers,showpacs,amsmath,amssymb]{revtex4}
\usepackage{graphicx}
\usepackage{amssymb}

\begin{document}

\title{Practical quantum oblivious transfer with a single photon}
\author{Guang Ping He}
\email{hegp@mail.sysu.edu.cn}
\affiliation{School of Physics, Sun Yat-sen University, Guangzhou 510275, China}

\begin{abstract}
Quantum oblivious transfer (QOT) is an essential cryptographic primitive.
But unconditionally secure QOT is known to be impossible. Here we propose a
practical QOT protocol, which is perfectly secure against dishonest sender
without relying on any technological assumption. Meanwhile, it is also
secure against dishonest receiver in the absence of long-term quantum memory
and complicated collective measurements. The protocol is extremely feasible,
as it can be implemented using currently available Mach-Zehnder
interferometer, and no quantum memory, collective measurements nor
entanglement are needed for honest participants. More importantly, comparing with
other practical QOT schemes, our protocol has an unbeatable efficiency since
it requires the transmission of a single photon only. %\\
%
%Keywords: Mach-Zehnder interferometer, Quantum cryptography, Oblivious transfer, Communication security
\end{abstract}


%\keywords{Mach-Zehnder interferometer \and Quantum cryptography \and Oblivious transfer \and Communication security}

\maketitle

%%%%%%%%%%%%%%%%%%%%%%%%%%%%%%%%%%%%%%%%%%%%%%%%%%%%%%%%%%%%%%%%%%%%%%%%%%%%%%%%%%%%%%%%%%%%%%%%%%%%%%%%%%%%%%%%%%%%%%%%%%%%%%%%%%%%%%%%%%%%%%%%%%%%%%%%%%%%%%%%%%%%%%%%%%%%%%%%%%%%%%%%%%%%%%%%%%%%%%%%%%%%%%%%%%%%%%%%%%%%%%%%%%%%%%%%%%%%%%%%%%%%%%%%%%%%

\section{Introduction}

Oblivious transfer (OT) is an important concept in cryptography, with many
variations. The original version, a.k.a. all-or-nothing OT, was introduced
by Rabin in 1981 \cite{qbc9}. Shortly after, Even, Goldreich and Lempel
proposed 1-out-of-2 OT \cite{1-2OT}. Later it was proven that OT is an
essential building block for two-party and multi-party protocols \cite{qi139}%
. Unfortunately, unconditionally secure OT (whose security is guaranteed
solely by the fundamental laws of physics) was proven impossible \cite%
{qi149,qi500,qi499,qi677,qi725,qi797,qbc14}. Even OT built-upon secure
relativistic bit commitment \cite{qi44,qi582,qbc24,qbc51} is covered too
\cite{HeEPJD}.

To circumvent the problem, many practically secure quantum OT (QOT) were
proposed, whose security relies on the assumption that the adversary is
limited to some technological constraints, e.g., noisy quantum storage\
(where the quantum memory is subjected to a certain level of imperfection)
\cite{qi601,qbc190,qi795,qbc6,qi796,qbc86,qbc154,qbc155} or bounded quantum
storage\ (where the amount of quantum data that can be stored is limited by
some constants) \cite{qi243,qbc39}. But to transfer a single classical bit
obliviously, these protocols have to involve the transmission of thousands
of qubits to reach a satisfactory level of security. Consequently, if they
are used for building more complicated multi-party secure computation
protocols \cite{qi139} which may call for many rounds of QOT, then the total
number of transmitted qubits will be tremendous, making the resultant
protocols lose the sense of being \textquotedblleft
practical\textquotedblright .

Here we will propose a QOT protocol, in which only the transmission of a
single photon is required. For dishonest sender, our protocol is not only
unconditionally secure (i.e., the cheating probability can be made
arbitrarily close to $0$ as long as the fundamental laws of physics hold),
but also perfectly secure (the sender's cheating probability equal to $0$\
exactly). Meanwhile, it is practically secure against dishonest receiver as
long as he does not have long-term quantum memory and cannot perform
complicated collective measurements.

Such a requirement on quantum memory is much the same to the above
noisy-storage model, except that we explicitly require\ that the
imperfection of the storage devices should reach the extent that after a
certain period of time, the final state in the quantum memory should no
longer contain any information about the original state that it initially
stored. This can easily be met nowadays, because in practice, due to the
unavoidable noise from the environment, no quantum memory can store a
quantum state free from error forever. With the help of quantum correction
code, the lifetime of the state can be extended. But implementing quantum
correction code is still hard for currently available technology. Thus our
protocol will be useful today and in the foreseen future.

\section{Definitions}

We will focus only on all-or-nothing OT below, so we simply call it OT. As defined in \cite{qi139}, OT is a cryptographic task
between sender Alice and receiver Bob, with the following properties.

\bigskip

\textbf{OT:}

(I) Alice has a secret bit $b$.

(II) At the end of the protocol, one of the following events occurs, each
with probability $1/2$: (1) Bob learns the value of $b$. (2) Bob gains no
information about $b$.

(III) Bob knows which of these two events actually occurred.

(IV) Alice learns nothing about whether Bob got $b$ or not.

\bigskip

From these definitions we can see that the goal of dishonest Alice is to
learn what happens at Bob's side, while dishonest Bob wants to increase his
probability of getting $b$. Let $P_{Alice}^{\ast }$ denote the probability
that Alice knows whether Bob got $b$ or not. Note that even if the protocol
tells Alice nothing about what happens to Bob, she can still make a random
guess, which can be correct with probability $1/2$,
%. Therefore, an OT protocol is considered secure against Alice as long as it %can guarantee
%\begin{equation}
%P_{Alice}^{\ast }=\frac{1}{2}.
%\end{equation}%
i.e., there is $%
P_{Alice}^{\ast }=1/2$ even in the honest case. Thus, \textbf{Alice's
successful cheating probability} $v$\ should be defined as%
\begin{equation}
v\equiv P_{Alice}^{\ast }-\frac{1}{2}.  \label{v}
\end{equation}

Note that forcing Alice to commit to a specific value of $b$ is not the duty
of OT. If she conceives the value of $b$ in her mind while her actual
actions in the protocol lead Bob to another irrelevant bit $\tilde{b}$, then
we should take $\tilde{b}$\ as the actual $b$ value that Alice inputs to the
protocol instead. This should not be considered as cheating, as long as it
does not increase Alice's probability $v$ on knowing whether Bob got $b$ or
not.

When studying Bob's cheating, we should note that Bob can choose
to either get $b$ unambiguously (i.e., he knows with certainty whether the
value he got is correct or not) or ambiguously (i.e., his obtained value
matches $b$\ with a considerable probability, but he does not known exactly
whether it is correct or not). Let $b^{\prime }$ denote what Bob obtained
from the protocol, we can define the \textbf{reliability} $R$\ of $b^{\prime
}$ as the probability for $b^{\prime }=b$. When Bob gets $b$ ambiguously
(unambiguously), there is $R<100\%$\ ($R=100\%$).

Let $P_{Bob}^{\ast }$ denote the probability that Bob gets $b$ with
reliability $R=100\%$. Then
%the definition of OT implies that
the goal of secure OT is to ensure
%\begin{equation}
%P_{Bob}^{\ast }=\frac{1}{2}.
%\end{equation}
 $P_{Bob}^{\ast }=1/2$. Therefore, it is natural to define \textbf{Bob's successful cheating
probability} $u$\ as%
\begin{equation}
u\equiv P_{Bob}^{\ast }-\frac{1}{2}.  \label{u}
\end{equation}%
Nevertheless, when Bob knows unambiguously that he fails to get $b$, he can
still make a guess, which can be correct with probability $1/2$, i.e., $%
R=50\%$. Thus we should note that the average reliability that Bob obtains in secure OT is not required to be
equal to $1/2$. Instead, it should be%
\begin{equation}
\bar{R}=\frac{1}{2}\times 100\%+\frac{1}{2}\times 50\%=75\%.  \label{R}
\end{equation}

\bigskip

%%%%%%%%%%%%%%%%%%%%%%%%%%%%%%%%%%%%%%%%%%%%%%%%%%%%%%%%%%%%%%%%%%%%%%%%

\begin{figure*}[tbp]
\includegraphics{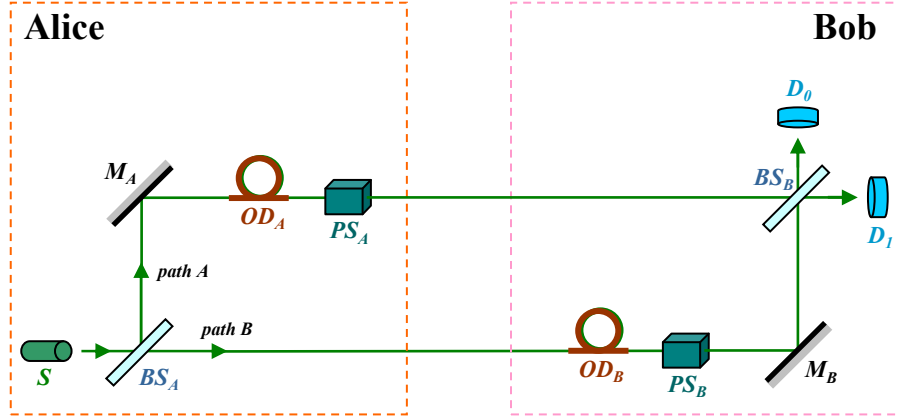}
\caption{Diagram of our OT protocol. Alice sends a single photon from the source $S$ to the beam
splitter $BS_{A}$, which is a half-silvered mirror that will either reflect the photon into path $A$\ or transmit it into path $B$ with equal
probabilities. On path $A$\ ($B$), the photon is reflected by the mirror $%
M_{A}$ ($M_{B}$) so that it is redirected to another half-reflected and
half-transmitted beam splitter $BS_{B}$ after passing the optical delay $%
OD_{A}$\ ($OD_{B}$) and the phase shifter $PS_{A}$\ ($PS_{B}$), then reaches
either the detector $D_{0}$ or $D_{1}$. When $OD_{A}$\ and $OD_{B}$ are set
to produce the same delay time, the total length of paths $A$\ and $B$\
are equal, so that the complete apparatus acts as a balanced Mach-Zehnder
interferometer.}
\label{fig:epsart}
\end{figure*}

%%%%%%%%%%%%%%%%%%%%%%%%%%%%%%%%%%%%%%%%%%%%%%%%%%%%%%%%%%%%%%%%%%%%%%%%

\section{Our protocol}

As mentioned in the introduction, in practice any quantum memory has a
limited storage time. Let $T$ denote the upperbound of the storage time of
state-of-the-art quantum memory (even when quantum correction is taken into
account), such that after time $T$, we can be sure that it is no longer
possible to gain any information of the state initially stored in the
memory. With this consideration, we propose a practical QOT protocol, whose
experimental apparatus is illustrated in Fig.1. Let $\tau _{A}$ ($\tau _{B}$%
) denote the delay time that Alice (Bob) introduced using $OD_{A}$\ ($OD_{B}$%
), and $\theta _{A}$ ($\theta _{B}$) be the phase shift angle produced by $%
PS_{A}$\ ($PS_{B}$). For simplicity, suppose that except for $OD_{A}$\ and $%
OD_{B}$, the time for the photon to travel through all other devices in Fig.1 is negligible. Our protocol is as follows.

\bigskip

\textbf{Protocol QOT:} (for transferring a bit $b\in \{0,1\}$ from Alice to
Bob)

(i) Alice and Bob agree on the times $t_{1}$ and $t_{2}$ which mark the
beginning and the end of the transmission, and a fixed delay time value $%
\Delta \ll t_{2}-t_{1}$.

(ii) Bob randomly picks the delay time $\tau _{B}\in \{0,\Delta \}$ and the
phase shift angle $\theta _{B}\in \{0,\pi \}$, and sets the optical delay $%
OD_{B}$\ and the phase shifter $PS_{B}$\ accordingly. He keeps them in these
settings during the entire time interval $[t_{1},t_{2}]$.

(iii) Alice picks the delay time $\tau _{A}\in \{0,\Delta \}$ randomly, and
takes the phase shift angle%
\begin{equation}
\theta _{A}=b\pi  \label{thetaA}
\end{equation}%
where $b\in \{0,1\}$ is the secret bit that she wants to transfer, then sets
$OD_{A}$\ and $PS_{A}$\ accordingly. She also picks a secret time $t_{s}\in
\lbrack t_{1},t_{2}-\Delta ]$ randomly, and sends a photon from the source $%
S $ at $t_{s}$.

(iv) When Bob finds his detector $D_{i}$ ($i\in \{0,1\}$) clicks, he records
the index $i$. Note that if he detects more than one click or no click at
all within $[t_{1},t_{2}]$, he should conclude that Alice cheats.

(v) After time $t_{2}+T$, Alice announces $\tau _{A}$ to Bob.

(v) If Bob finds $\tau _{A}\neq \tau _{B}$, he concludes that he fail to get
$b$. Or if $\tau _{A}=\tau _{B}$, he concludes that%
\begin{equation}
b=i\oplus (\theta _{B}/\pi ).  \label{b}
\end{equation}

\bigskip

\section{Correctness}

Correctness means that if both parties are honest, then the goal of OT can
be reached. In our protocol, when using the second quantization formalism,
we can use $\left\vert t\right\rangle _{A}\left\vert 0\right\rangle _{B}$\
to denote that there is a photon on path $A$ at time $t$\ and no photon on
path $B$, and use $\left\vert 0\right\rangle _{A}\left\vert t\right\rangle
_{B}$\ to denote that there is a photon on path $B$ at time $t$ and no
photon on path $A$. Then the initial state of Alice's photon after passing $%
BS_{A}$\ is%
\begin{equation}
\left\vert \psi \right\rangle _{in}=\frac{1}{\sqrt{2}}(\left\vert
t_{s}\right\rangle _{A}\left\vert 0\right\rangle _{B}+\left\vert
0\right\rangle _{A}\left\vert t_{s}\right\rangle _{B}).  \label{initial}
\end{equation}%
After passing $OD_{A}$, $PS_{A}$, $OD_{B}$ and $PS_{B}$, the final state of
the photon arriving at $BS_{B}$\ is%
\begin{equation}
\left\vert \psi \right\rangle _{f}=\frac{1}{\sqrt{2}}(e^{i\theta
_{A}}\left\vert t_{s}+\tau _{A}\right\rangle _{A}\left\vert 0\right\rangle
_{B}+e^{i\theta _{B}}\left\vert 0\right\rangle _{A}\left\vert t_{s}+\tau
_{B}\right\rangle _{B}).  \label{final}
\end{equation}

Meanwhile, when combining with $BS_{B}$, $D_{0}$ and $D_{1}$ serve as the
projective operators%
\begin{equation}
P_{0}\equiv \left\vert \psi \right\rangle _{0}\left\langle \psi \right\vert
_{0}  \label{p0}
\end{equation}%
and%
\begin{equation}
P_{1}\equiv \left\vert \psi \right\rangle _{1}\left\langle \psi \right\vert
_{1},  \label{p1}
\end{equation}%
respectively, where%
\begin{equation}
\left\vert \psi \right\rangle _{0}\equiv \frac{1}{\sqrt{2}}(\left\vert
t\right\rangle _{A}\left\vert 0\right\rangle _{B}+\left\vert 0\right\rangle
_{A}\left\vert t\right\rangle _{B}),
\end{equation}%
and%
\begin{equation}
\left\vert \psi \right\rangle _{1}\equiv \frac{1}{\sqrt{2}}(\left\vert
t\right\rangle _{A}\left\vert 0\right\rangle _{B}-\left\vert 0\right\rangle
_{A}\left\vert t\right\rangle _{B}).
\end{equation}

Since $\tau _{A},\tau _{B}\in \{0,\Delta \}$, with probability $1/2$ there
is $\tau _{A}=\tau _{B}$. Then if $\theta _{A}=\theta _{B}$, Eq.(\ref{final}%
) becomes%
\begin{equation}
\left\vert \psi \right\rangle _{f}=e^{i\theta _{A}}\frac{1}{\sqrt{2}}%
(\left\vert t_{s}+\tau _{A}\right\rangle _{A}\left\vert 0\right\rangle
_{B}+\left\vert 0\right\rangle _{A}\left\vert t_{s}+\tau _{A}\right\rangle
_{B}),
\end{equation}%
so that the photon will be detected by $D_{0}$ at time $t_{s}+\tau _{A}$
with certainty. Or if $\left\vert \theta _{A}-\theta _{B}\right\vert =\pi $,
then%
\begin{equation}
\left\vert \psi \right\rangle _{f}=e^{i\theta _{A}}\frac{1}{\sqrt{2}}%
(\left\vert t_{s}+\tau _{A}\right\rangle _{A}\left\vert 0\right\rangle
_{B}-\left\vert 0\right\rangle _{A}\left\vert t_{s}+\tau _{A}\right\rangle
_{B}),
\end{equation}%
so that the photon will be detected by $D_{1}$ at time $t_{s}+\tau _{A}$
with certainty. That is, in either case the detector $D_{i}$\ will click,
where%
\begin{equation}
i=(\theta _{A}/\pi )\oplus (\theta _{B}/\pi ).  \label{i}
\end{equation}%
Combining with Eq.(\ref{thetaA}), we can see that the value of $b$ that Bob
obtained from Eq.(\ref{b}) is correct.

On the other hand, with probability $1/2$ there will be $\tau _{A}\neq \tau
_{B}$. Then Eq.(\ref{final}) indicates that the state of the photon arrived
at $BS_{B}$\ at time $t_{s}+\tau _{A}$\ (or $t_{s}+\tau _{B}$) has the form $%
\left\vert t_{s}+\tau _{A}\right\rangle _{A}\left\vert 0\right\rangle _{B}$
(or $\left\vert 0\right\rangle _{A}\left\vert t_{s}+\tau _{B}\right\rangle
_{B}$). Applying Eqs.(\ref{p0}) and (\ref{p1}) on them, we know that $D_{0}$
and $D_{1}$ could click with equal probabilities at either time $t_{s}+\tau
_{A}$\ or $t_{s}+\tau _{B}$. Therefore, unlike Eq.(\ref{i}), now the index $%
i $ of the detector $D_{i}$\ that clicks contains no information on the
relationship between $\theta _{A}$ and $\theta _{B}$, so that Bob fails to
get $b$.

Thus we proved that when both parties are honest, our protocol can ensure
that with probability $1/2$, Bob can get $b$ with reliability $R=100\%$.

\bigskip

\section{Security against Alice}

The goal of dishonest Alice is to know whether Bob got $b$ or not. According
to step(v), Bob got $b$ (failed to get $b$) if $\tau _{A}=\tau _{B}$ ($\tau
_{A}\neq \tau _{B}$). Thus, Alice needs to know Bob's choice of $\tau _{B}$.
But a distinct feature of our protocol is: there is only one-way information
transfer from Alice to Bob, while Bob never announces any classical
information or transfers any quantum information to Alice. Thus it is clear
that Alice has no method to learn what happens at Bob's side. All that she
can do is to make a random guess, which can be correct with probability
%\begin{equation}
%P_{Alice}^{\ast }=\frac{1}{2}.
%\end{equation}%
$P_{Alice}^{\ast }=1/2$. Thus her
successful cheating probability is exactly
\begin{equation}
v=P_{Alice}^{\ast }-\frac{1}{2}=0.  \label{v2}
\end{equation}
Therefore, our protocol is perfectly secure against dishonest Alice, without relying on any technological assumption.

\bigskip

\section{Security against Bob}

In principle, if Bob can store the photon unmeasured until Alice announces $%
\tau _{A}$, then he can always choose $\tau _{A}=\tau _{B}$ and learn $b$
with probability $100\%$. Thus our protocol is not unconditionally secure.
But in practice, as we assumed that the maximum storage time of the state of
the photon is $T$, Bob has to complete the measurement no later then $t_{2}+T
$. At that time Alice has not announces $\tau _{A}$ yet. This forces Bob to
make his own choice of $\tau _{B}$ without knowing $\tau _{A}$.
Consequently, it guarantees that $\tau _{A}=\tau _{B}$ occurs with
probability $1/2$. When this indeed occurs, Bob gets $b$ with reliability $%
R=100\%$. Else if $\tau _{A}\neq \tau _{B}$ occurs, Bob knows that he fails
to get $b$. Thus the probability that Bob gets $b$ with reliability $R=100\%$ is
%\begin{equation}
%P_{Bob}^{\ast }=\frac{1}{2},
%\end{equation}
$P_{Bob}^{\ast }=1/2$, i.e., his cheating probability is
\begin{equation}
u=P_{Bob}^{\ast }-\frac{1}{2}=0  \label{u2}
\end{equation}%
so that our protocol meets the requirement of OT in practice.

However, there is still a security problem to worry. Let $\rho _{0}$
and $\rho _{1}$ denote the density matrices of the unmeasured state of the
photon entering Bob's site (before passing $OD_{B}$ and $PS_{B}$)
corresponding to Alice's choices $b=0$ and $b=1$, respectively. As
elaborated in the appendix, they can be written as
\begin{eqnarray}
\rho _{0} &=&\frac{1}{2(2n-\Delta )}(\sum_{t_{s}=t_{1}}^{t_{n}}((\left\vert
t_{s}\right\rangle _{A}\left\vert 0\right\rangle _{B}+\left\vert
0\right\rangle _{A}\left\vert t_{s}\right\rangle _{B})  \nonumber \\
&&\times (\left\langle t_{s}\right\vert _{A}\left\langle 0\right\vert
_{B}+\left\langle 0\right\vert _{A}\left\langle t_{s}\right\vert _{B}))
\nonumber \\
&&+\sum_{t_{s}=t_{1}}^{t_{n-\Delta }}((\left\vert t_{s}+\Delta \right\rangle
_{A}\left\vert 0\right\rangle _{B}+\left\vert 0\right\rangle _{A}\left\vert
t_{s}\right\rangle _{B})  \nonumber \\
&&\times (\left\langle t_{s}+\Delta \right\vert _{A}\left\langle
0\right\vert _{B}+\left\langle 0\right\vert _{A}\left\langle
t_{s}\right\vert _{B})))  \label{rou0}
\end{eqnarray}%
and
\begin{eqnarray}
\rho _{1} &=&\frac{1}{2(2n-\Delta )}(\sum_{t_{s}=t_{1}}^{t_{n}}((-\left\vert
t_{s}\right\rangle _{A}\left\vert 0\right\rangle _{B}+\left\vert
0\right\rangle _{A}\left\vert t_{s}\right\rangle _{B})  \nonumber \\
&&\times (-\left\langle t_{s}\right\vert _{A}\left\langle 0\right\vert
_{B}+\left\langle 0\right\vert _{A}\left\langle t_{s}\right\vert _{B}))
\nonumber \\
&&+\sum_{t_{s}=t_{1}}^{t_{n-\Delta }}((-\left\vert t_{s}+\Delta
\right\rangle _{A}\left\vert 0\right\rangle _{B}+\left\vert 0\right\rangle
_{A}\left\vert t_{s}\right\rangle _{B})  \nonumber \\
&&\times (-\left\langle t_{s}+\Delta \right\vert _{A}\left\langle
0\right\vert _{B}+\left\langle 0\right\vert _{A}\left\langle
t_{s}\right\vert _{B}))),  \label{rou1}
\end{eqnarray}%
where we treat the time interval $[t_{1},t_{2}]$\ as $n$ equally divided
time slots, i.e., $t_{s}$ can be chosen as $t_{1}$, $t_{2}$, $...$, $t_{n}$,
with $n\rightarrow \infty $. Then numerical calculation shows that the trace
distance between $\rho _{0}$ and $\rho _{1}$ for high $n$\ is%
\begin{equation}
D(\rho _{0},\rho _{1})\equiv \frac{1}{2}tr\sqrt{(\rho _{0}-\rho _{1})^{\dag
}(\rho _{0}-\rho _{1})}\simeq 0.637.
\end{equation}%
According to Theorem 9.1 of \cite{qi366}, this indicates that there could be
a positive-operator valued measure (POVM) for Bob, which can discriminate $%
\rho _{0}$ and $\rho _{1}$ (and thus learn $b$) unambiguously with
probability%
\begin{equation}
p=D(\rho _{0},\rho _{1}),
\end{equation}%
or can discriminate $\rho _{0}$ and $\rho _{1}$ ambiguously so that Bob get $%
b$ with the average reliability%
\begin{equation}
\bar{R}=(D(\rho _{0},\rho _{1})+1)/2.
\end{equation}%
Consequently, if Bob can apply this POVM, he can cheat with probability
\begin{equation}
u=p-1/2\simeq 0.137,  \label{u3}
\end{equation}%
or obtain $b$ with an average reliability%
\begin{equation}
\bar{R}\simeq 81.9\%>75\%
\end{equation}
so that our protocol is not a secure OT in principle.

Nevertheless, currently it is unknown whether such a POVM can be implemented
with currently available technology or not. Especially, Eqs.(\ref{rou0}) and (%
\ref{rou1}) show that $\rho _{0}$ and $\rho _{1}$ contain the states from
different time $t_{s}$ where $s=1,2,...,n$ with $n\rightarrow \infty $. Thus
the POVM for discriminating $\rho _{0}$ and $\rho _{1}$ is actually a
collective measurement on an infinite number of states potentially being
sent at different time $t_{s}$. Also, in practice the error rate of quantum
memory increases with time. Then the actual density matrices $\rho _{0}$ and
$\rho _{1}$ will then turn into the mixes of quantum states containing different amount of error. It makes the POVM for decoding the correct $b$ become even more complicated. Therefore,
before these difficulties could be overcome in the future, our protocol can
be trusted as a secure OT in practice.

\section{Discussions}

In summary, we propose an OT protocol, which is perfectly secure against
dishonest Alice, i.e., her probability on learning whether Bob got $b$ or
not is exactly zero. Meanwhile, it is practically secure against dishonest
Bob, i.e., his probability for getting $b$ with reliability $100\%$ is
exactly $1/2$ as long as he does not have long-term quantum memory nor the
POVM for maximally discriminating $\rho _{0}$ and $\rho _{1}$ in Eqs.(\ref%
{rou0}) and (\ref{rou1}) without knowing Alice's delay time $\tau _{A}$.

%Comparing with the weak OT protocol in Ref.\cite{qbc20}, our protocol has
%better security. When taking{\large \ }$a=\cos \alpha =1/2$ in the protocol
%in Ref.\cite{qbc20}, we can achieve the goal that honest Bob can get
%Alice's bit unambiguously with probability $p=1/2$. But in this case,
%according to section 2.3 of Ref.\cite{qbc20}, Alice can cheat with
%probability $v=a(1-a)/(1+a)=1/6$. Meanwhile, Bob's maximum probability of
%getting Alice's bit (even ambiguously) is $q=\cos ^{2}\theta =\cos ^{2}(\pi
%/4-\alpha /2)\simeq 0.933$. Both probabilities are higher than these of ours.

On the feasibility aspect, by comparing our Fig.1 and that of \cite{qi889},
we can see that the technology in \cite{qi889} is sufficient for
implementing our protocol. Also, honest participants do not need quantum
memory, collective measurements nor entanglement. Thus it is very feasible.
Comparing with other practical OT protocols using noisy or bounded storage
\cite{qi601,qbc190,qi795,qbc6,qi796,qbc86,qbc154,qbc155,qi243,qbc39}, our protocol only needs the
transmission of a single photon. Thus the efficiency is unbeatable.

In the above analysis, we have not considered the transmission and detection
errors in practice. But since only one photon is used in our protocol, it is
obvious that the error rate of the final transferred bit $b$ is determined
directly by the error rate in transmission and detection, which can easily
be checked in practice. %\\

%\section*{Acknowledgements}

%This research did not receive any specific grant from funding agencies in the public, commercial, or not-for-profit sectors.
%The work was supported in part by the National Science Foundation of
%Guangdong province.
%China.

\section*{Appendix: The density matrices and trace distance}

Here we calculate the density matrices $\rho _{0}$, $\rho _{1}$ and trace
distance $D(\rho _{0},\rho _{1})$ used in the security proof of our protocol
against dishonest Bob.

After passing $OD_{A}$ and $PS_{A}$, the state of the photon that Alice sent
to Bob (i.e., $\left\vert \psi \right\rangle _{in}$ in Eq.(\ref{initial})) turns into
\begin{equation}
\left\vert \psi \right\rangle _{s}=\frac{1}{\sqrt{2}}(e^{i\theta
_{A}}\left\vert t_{s}+\tau _{A}\right\rangle _{A}\left\vert 0\right\rangle
_{B}+\left\vert 0\right\rangle _{A}\left\vert t_{s}\right\rangle _{B}).
\end{equation}%
That is, if $b=0$, then
\begin{equation}
\left\vert \psi \right\rangle _{s}=\frac{1}{\sqrt{2}}(\left\vert t_{s}+\tau
_{A}\right\rangle _{A}\left\vert 0\right\rangle _{B}+\left\vert
0\right\rangle _{A}\left\vert t_{s}\right\rangle _{B}),
\end{equation}%
or if $b=1$, then
\begin{equation}
\left\vert \psi \right\rangle _{s}=\frac{1}{\sqrt{2}}(-\left\vert t_{s}+\tau
_{A}\right\rangle _{A}\left\vert 0\right\rangle _{B}+\left\vert
0\right\rangle _{A}\left\vert t_{s}\right\rangle _{B}).
\end{equation}%
Let us treat the time interval $[t_{1},t_{2}]$\ as $n$ equally divided time
slots with $n\rightarrow \infty $, i.e., $t_{s}$ can be chosen as $t_{1}$, $%
t_{2}$, $...$, $t_{n}$. Since $\tau _{A}$\ can be either $0$ or $\Delta $
with equal probabilities, the density matrices corresponding to $b=0$ and $%
b=1$, respectively, are
\begin{eqnarray}
\rho _{0} &=&\frac{1}{2(2n-\Delta )}(\sum_{t_{s}=t_{1}}^{t_{n}}((\left\vert
t_{s}\right\rangle _{A}\left\vert 0\right\rangle _{B}+\left\vert
0\right\rangle _{A}\left\vert t_{s}\right\rangle _{B})  \nonumber \\
&&\times (\left\langle t_{s}\right\vert _{A}\left\langle 0\right\vert
_{B}+\left\langle 0\right\vert _{A}\left\langle t_{s}\right\vert _{B}))
\nonumber \\
&&+\sum_{t_{s}=t_{1}}^{t_{n-\Delta }}((\left\vert t_{s}+\Delta \right\rangle
_{A}\left\vert 0\right\rangle _{B}+\left\vert 0\right\rangle _{A}\left\vert
t_{s}\right\rangle _{B})  \nonumber \\
&&\times (\left\langle t_{s}+\Delta \right\vert _{A}\left\langle
0\right\vert _{B}+\left\langle 0\right\vert _{A}\left\langle
t_{s}\right\vert _{B})))
\end{eqnarray}%
and
\begin{eqnarray}
\rho _{1} &=&\frac{1}{2(2n-\Delta )}(\sum_{t_{s}=t_{1}}^{t_{n}}((-\left\vert
t_{s}\right\rangle _{A}\left\vert 0\right\rangle _{B}+\left\vert
0\right\rangle _{A}\left\vert t_{s}\right\rangle _{B})  \nonumber \\
&&\times (-\left\langle t_{s}\right\vert _{A}\left\langle 0\right\vert
_{B}+\left\langle 0\right\vert _{A}\left\langle t_{s}\right\vert _{B}))
\nonumber \\
&&+\sum_{t_{s}=t_{1}}^{t_{n-\Delta }}((-\left\vert t_{s}+\Delta
\right\rangle _{A}\left\vert 0\right\rangle _{B}+\left\vert 0\right\rangle
_{A}\left\vert t_{s}\right\rangle _{B})  \nonumber \\
&&\times (-\left\langle t_{s}+\Delta \right\vert _{A}\left\langle
0\right\vert _{B}+\left\langle 0\right\vert _{A}\left\langle
t_{s}\right\vert _{B}))),
\end{eqnarray}%
thus Eqs.(\ref{rou0}) and (\ref{rou1}) in the main text are obtained.

Now let us calculate the trace distance between them. The above equations
give
\begin{eqnarray}
\rho _{0}-\rho _{1} &=&\frac{1}{2n-\Delta }(\sum_{t_{s}=t_{1}}^{t_{n}}(\left%
\vert t_{s}\right\rangle _{A}\left\vert 0\right\rangle _{B}\left\langle
0\right\vert _{A}\left\langle t_{s}\right\vert _{B}  \nonumber \\
&&+\left\vert 0\right\rangle _{A}\left\vert t_{s}\right\rangle
_{B}\left\langle t_{s}\right\vert _{A}\left\langle 0\right\vert _{B})
\nonumber \\
&&+\sum_{t_{s}=t_{1}}^{t_{n-\Delta }}(\left\vert t_{s}+\Delta \right\rangle
_{A}\left\vert 0\right\rangle _{B}\left\langle 0\right\vert _{A}\left\langle
t_{s}\right\vert _{B}  \nonumber \\
&&+\left\vert 0\right\rangle _{A}\left\vert t_{s}\right\rangle
_{B}\left\langle t_{s}+\Delta \right\vert _{A}\left\langle 0\right\vert
_{B})).
\end{eqnarray}%
Denote%
\begin{eqnarray}
\left\vert s\right\rangle  &\equiv &\left\vert t_{s}\right\rangle
_{A}\left\vert 0\right\rangle _{B}, \\
\left\vert s+n\right\rangle  &\equiv &\left\vert 0\right\rangle
_{A}\left\vert t_{s}\right\rangle _{B}
\end{eqnarray}%
for $s=1,...,n$, and take $\Delta =1$ for simplicity, then
\begin{eqnarray}
\rho _{0}-\rho _{1} &=&\frac{1}{2n-1}(\sum_{s=1}^{n}(\left\vert
s\right\rangle \left\langle s+n\right\vert +\left\vert s+n\right\rangle
\left\langle s\right\vert )  \nonumber \\
&&+\sum_{s=1}^{n-1}(\left\vert s+1\right\rangle \left\langle s+n\right\vert
+\left\vert s+n\right\rangle \left\langle s+1\right\vert )).  \nonumber \\
&&
\end{eqnarray}%
For $s=1,...,2n$, denote%
\begin{equation}
\left\langle s\right\vert =[%
\begin{array}{ccccccc}
0 & ... & 0 & 1 & 0 & ... & 0%
\end{array}%
]
\end{equation}%
i.e., only the $s$th element of the vector at the right-hand side of the equation is $1$, while the rest elements are all $0$.
Then
\begin{eqnarray}
\rho _{0}-\rho _{1} &=&\frac{1}{2n-1}(\sum_{s=1}^{n}(\left[ 1_{s,s+n}\right]
+\left[ 1_{s+n,s}\right] )  \nonumber \\
&&+\sum_{s=1}^{n-1}(\left[ 1_{s+1,s+n}\right] +\left[ 1_{s+n,s+1}\right] )),
\end{eqnarray}%
where $\left[ 1_{i,j}\right] $\ denote a $2n\times 2n$ matrix whose elements
are all $0$ except that the element in the $i$ row and $j$\ column is $1$.
Therefore
\begin{eqnarray}
\rho _{0}-\rho _{1} &=&\frac{1}{2n-1}  \nonumber \\
&&\times \left[
\begin{array}{cccccccccccccc}
0 & 0 & 0 & \cdots  & 0 & 0 & 0 & 1 & 0 & 0 & \cdots  & 0 & 0 & 0 \\
0 & 0 & 0 & \cdots  & 0 & 0 & 0 & 1 & 1 & 0 & \cdots  & 0 & 0 & 0 \\
0 & 0 & 0 & \cdots  & 0 & 0 & 0 & 0 & 1 & 1 & \cdots  & 0 & 0 & 0 \\
\vdots  & \vdots  & \vdots  & \ddots  & \vdots  & \vdots  & \vdots  & \vdots
& \vdots  & \vdots  & \vdots  & \vdots  & \vdots  & \vdots  \\
0 & 0 & 0 & \cdots  & 0 & 0 & 0 & 0 & 0 & 0 & \cdots  & 1 & 0 & 0 \\
0 & 0 & 0 & \cdots  & 0 & 0 & 0 & 0 & 0 & 0 & \cdots  & 1 & 1 & 0 \\
0 & 0 & 0 & \cdots  & 0 & 0 & 0 & 0 & 0 & 0 & \cdots  & 0 & 1 & 1 \\
1 & 1 & 0 & \cdots  & 0 & 0 & 0 & 0 & 0 & 0 & \cdots  & 0 & 0 & 0 \\
0 & 1 & 1 & \cdots  & 0 & 0 & 0 & 0 & 0 & 0 & \cdots  & 0 & 0 & 0 \\
0 & 0 & 1 & \cdots  & 0 & 0 & 0 & 0 & 0 & 0 & \cdots  & 0 & 0 & 0 \\
\vdots  & \vdots  & \vdots  & \vdots  & \vdots  & \vdots  & \vdots  & \vdots
& \vdots  & \vdots  & \ddots  & \vdots  & \vdots  & \vdots  \\
0 & 0 & 0 & \cdots  & 1 & 1 & 0 & 0 & 0 & 0 & \cdots  & 0 & 0 & 0 \\
0 & 0 & 0 & \cdots  & 0 & 1 & 1 & 0 & 0 & 0 & \cdots  & 0 & 0 & 0 \\
0 & 0 & 0 & \cdots  & 0 & 0 & 1 & 0 & 0 & 0 & \cdots  & 0 & 0 & 0%
\end{array}%
\right] ,  \nonumber \\
&&
\end{eqnarray}%
and we have
\begin{eqnarray}
(\rho _{0}-\rho _{1})^{\dag }(\rho _{0}-\rho _{1}) =\frac{1}{(2n-1)^{2}} \nonumber \\
\times \left[
\begin{array}{cccccccccccccc}
1 & 1 & 0 & \cdots  & 0 & 0 & 0 & 0 & 0 & 0 & \cdots  & 0 & 0 & 0 \\
1 & 2 & 1 & \cdots  & 0 & 0 & 0 & 0 & 0 & 0 & \cdots  & 0 & 0 & 0 \\
0 & 1 & 2 & \cdots  & 0 & 0 & 0 & 0 & 0 & 0 & \cdots  & 0 & 0 & 0 \\
\vdots  & \vdots  & \vdots  & \ddots  & \vdots  & \vdots  & \vdots  & \vdots
& \vdots  & \vdots  & \vdots  & \vdots  & \vdots  & \vdots  \\
0 & 0 & 0 & \cdots  & 2 & 1 & 0 & 0 & 0 & 0 & \cdots  & 0 & 0 & 0 \\
0 & 0 & 0 & \cdots  & 1 & 2 & 1 & 0 & 0 & 0 & \cdots  & 0 & 0 & 0 \\
0 & 0 & 0 & \cdots  & 0 & 1 & 2 & 0 & 0 & 0 & \cdots  & 0 & 0 & 0 \\
0 & 0 & 0 & \cdots  & 0 & 0 & 0 & 2 & 1 & 0 & \cdots  & 0 & 0 & 0 \\
0 & 0 & 0 & \cdots  & 0 & 0 & 0 & 1 & 2 & 1 & \cdots  & 0 & 0 & 0 \\
0 & 0 & 0 & \cdots  & 0 & 0 & 0 & 0 & 1 & 2 & \cdots  & 0 & 0 & 0 \\
\vdots  & \vdots  & \vdots  & \vdots  & \vdots  & \vdots  & \vdots  & \vdots
& \vdots  & \vdots  & \ddots  & \vdots  & \vdots  & \vdots  \\
0 & 0 & 0 & \cdots  & 0 & 0 & 0 & 0 & 0 & 0 & \cdots  & 2 & 1 & 0 \\
0 & 0 & 0 & \cdots  & 0 & 0 & 0 & 0 & 0 & 0 & \cdots  & 1 & 2 & 1 \\
0 & 0 & 0 & \cdots  & 0 & 0 & 0 & 0 & 0 & 0 & \cdots  & 0 & 1 & 1%
\end{array}%
\right] .  \nonumber \\
&&
\end{eqnarray}%
The trace distance between $\rho _{0}$ and $\rho _{1}$ is defined as%
\begin{equation}
D(\rho _{0},\rho _{1})\equiv \frac{1}{2}tr\sqrt{(\rho _{0}-\rho _{1})^{\dag
}(\rho _{0}-\rho _{1})}.
\end{equation}%
Numerical calculation shows that it drops as $n$ increases, but slows down
significantly when $n>50$, as shown in Fig.2. We calculated up
to $n=1000$, and the result is%
\begin{equation}
D(\rho _{0},\rho _{1})\simeq 0.637.
\end{equation}

%%%%%%%%%%%%%%%%%%%%%%%%%%%%%%%%%%%%%%%%%%%%%%%%%%%%%%%%%%%%%%%%%%%%%%%%

\begin{figure}[tbp]
\includegraphics{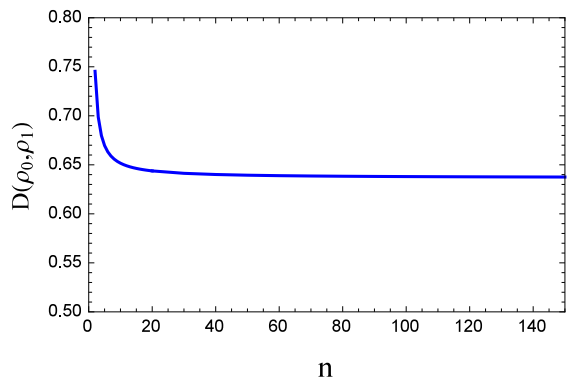}
\caption{The trace distance between the density matrice $\rho _{0}$ and $\rho _{1}$ in Eqs.(\ref{rou0}) and (\ref{rou1}) as a function of the number of the time slots $n$.}
\label{fig:epsart}
\end{figure}

%%%%%%%%%%%%%%%%%%%%%%%%%%%%%%%%%%%%%%%%%%%%%%%%%%%%%%%%%%%%%%%%%%%%%%%%

\end{document}